
\newif\ifPubSub \PubSubfalse
\newcount\PubSubMag \PubSubMag=1200
\def\PubSub{\PubSubtrue
            \magnification=\PubSubMag \hoffset=0pt \voffset=0pt
            \pretolerance=600 \tolerance=1200 \vbadness=1000 \hfuzz=2 true pt
            \baselineskip=1.75\baselineskip plus.2pt minus.1pt
            \parskip=2pt plus 6pt \gapskip=1.5\baselineskip
            \setbox\strutbox=\hbox{\vrule height .75\baselineskip
                                               depth  .25\baselineskip
                                               width 0pt}%
            \Page{5.75 true in}{8.9 true in}}

\newcount\FigNo \FigNo=0
\newbox\CapBox
\newbox\FigBox
\newtoks\0
\newif\ifImbed \Imbedfalse
\def\Fig {fig.~\the\FigNo}
\def\NFig {{\count255=\FigNo \advance\count255 by 1
            fig.\thinspace\the\count255}}

\def\StartFigure #1#2#3{\global\advance\FigNo by 1
                  \ifPubSub \global\setbox\FigBox=\vbox\bgroup
                            \unvbox\FigBox
                            \parindent=0pt \parskip0pt
                            \eject \line{\hfil}\bigskip\bigskip
                  \else \midinsert \removelastskip \vskip2\bigskipamount
                  \fi
                  \begingroup \hfuzz1in
                  \dimen0=\hsize \advance\dimen0 by-2\parindent \indent
                  \vbox\bgroup\hsize=\dimen0 \parindent=0pt \parskip=0pt
                       \vrule height1pt depth0pt width0pt
                       \ifdim\dimen0<#1bp \dimen1=#1bp
                                          \advance\dimen1 by -\dimen0
                                          \divide\dimen1 by 2 \hskip-\dimen1
                       \fi
                       \hfil
                       \vbox to #2 bp\bgroup\hsize #1 bp
                            \vss\noindent\strut\special{"#3}%
                            \skip0=\parskip \advance\skip0 by\dp\strutbox
                            \vskip-\skip0 }%
\def\caption #1{\strut\egroup
                \ifPubSub \global\setbox\CapBox=\vbox{\unvbox\CapBox
                                        \parindent0pt \medskip
                                         {\bf Figure \the\FigNo:}
                                          {\tenrm\strut #1\strut}}%
                \else \bigskip {\FootCapFace \hfuzz=1pt \baselineskip=3ex
                                \noindent{\bf Figure~\the\FigNo:} #1\par}%
                \fi}%
\def\label (#1,#2)#3{{\offinterlineskip \parindent=0pt \parskip=0pt
                     \vskip-\parskip 
                     \vbox to 0pt{\vss
                          \moveright #1bp\hbox to 0pt{\raise #2bp
                                         \hbox{#3}\hss}\hskip-#1bp\relax}}}%
\def\EndFigure {\egroup \endgroup 
                \ifPubSub \vfil
                          \centerline{\bf Figure \the\FigNo}
                          \egroup 
                \else \bigskip \endinsert 
                \fi \Imbedfalse}
\def\ListCaptions {\vfil\eject \message{!Figure captions:!}%
                   \Sectionvar{Figure Captions}\par
                   \unvbox\CapBox}
\def\ShowFigures {\ifPubSub\ifnum\FigNo>0 \ListCaptions \vfil\eject
                                 \message{!Figures:!}%
                                 \nopagenumbers \unvbox\FigBox \eject
                           \fi
                  \fi}
\def\FootCapFace{} 


\parskip=0pt plus 3pt
\def\Page #1#2{{\dimen0=\hsize \advance\dimen0 by-#1 \divide\dimen0 by 2
               \global\advance\hoffset by \dimen0
               \dimen0=\vsize \advance\dimen0 by-#2 \divide\dimen0 by 4
               \ifdim\dimen0<0pt \multiply\dimen0 by 3 \fi
               \global\advance\voffset by \dimen0
               \global\hsize=#1 \global\vsize=#2\relax}
               \ifdim\hsize<5.5in \tolerance=300 \pretolerance=300 \fi}
\def\EndPaper{\par\dosupereject \ifnum\RefNo>1 \ShowReferences \fi
              \ShowFigures
              \par\vfill\supereject
              \message{!!That's it.}\end}
\headline{\ifnum\pageno=-1 \hfil \Smallrm \Time, \Date \else \hfil \fi}
\def\VersionInfo #1{\headline{\ifnum\pageno=1 \hfil \Smallrm #1
                              \else \hfil
                              \fi}}


\newcount\RefNo \RefNo=1
\newbox\RefBox
\def\Jou #1{{\it #1}}
\def\Vol #1{{\bf #1}}
\def\AddRef #1{\setbox\RefBox=\vbox{\unvbox\RefBox
                       \parindent1.75em
                       \pretolerance=1000 \hbadness=1000
                       \vskip0pt plus1pt
                       \item{\the\RefNo.}
                       \sfcode`\.=1000 \strut#1\strut}%
               \global\advance\RefNo by1 }
\def\Ref  #1{\Hunskip~[{\the\RefNo}]\AddRef{#1}}
\def\Refc #1{\Hunskip~[{\the\RefNo,$\,$}\AddRef{#1}}
\def\Refm #1{\Hunskip{\the\RefNo,$\,$}\AddRef{#1}}
\def\Refe #1{\Hunskip{\the\RefNo}]\AddRef{#1}}
\def\Refl #1{\Hunskip~[{\the\RefNo}--\nobreak\AddRef{#1}}
\def\Refn #1{\Hunskip\AddRef{#1}}
\def\ShowReferences {\message{!References:!}%
                     \Sectionvar{References}\par
                     \vskip-5\parskip
                     \unvbox\RefBox}
\def\StoreRef #1{\Hunskip\edef#1{\the\RefNo}}


\newif\ifRomanNum
\newcount \Sno \Sno=0
\def\Interskip #1#2#3{{\removelastskip \dimen0=#1
                       \advance\dimen0 by2\baselineskip
                       \vskip0pt plus\dimen0 \penalty-300
                       \vskip0pt plus-\dimen0
                       \advance\dimen0 by -2\baselineskip
                       \vskip\dimen0 plus#2 minus#3}}
\def\Section #1\par{\Interskip{24pt}{6pt}{2pt}%
                    \global\advance\Sno by1
                    \setbox0=\hbox{\STitlefont
                                    \ifRomanNum
                                       \global\SSno=64 \uppercase
                                       \expandafter{\romannumeral
                                                    \the\Sno.\ \ }%
                                     \else
                                        \global\SSno=96 \the\Sno.\ \
                                     \fi}
                    \leftline{\vtop{\copy0 }%
                              \vtop{\advance\hsize by -\wd0
                                    \raggedright
                                    \pretolerance10000 \hbadness10000
                                    \noindent \STitlefont
                                    \GetParenDim{\dimen0}{\dimen1}%
                                    \advance\dimen0 by\dimen1
                                    \baselineskip=\dimen0 #1}}
                    \hrule height0pt depth0pt
                    \dimen0=\baselineskip \advance\dimen0 by -\parskip
                    \nobreak\vskip\dimen0 plus 3pt minus3pt \noindent}%
\def\Sectionvar #1\par{\Interskip{24pt}{6pt}{2pt}%
                    \leftline{\vbox{\noindent
                                    \STitlefont #1}}
                    \hrule height0pt depth0pt
                    \dimen0=\baselineskip \advance\dimen0 by -\parskip
                    \nobreak\vskip\dimen0 plus 3pt minus3pt \noindent}%
\newcount\SSno \SSno=0
\def\SubSection #1\par{\Interskip{15pt}{3pt}{1pt}%
                    \global\advance\SSno by1
                    \setbox0=\hbox{\SSTitlefont \char\SSno$\,$]$\,$\ }
                    \leftline{\vtop{\copy0 }%
                              \vtop{\advance\hsize by -\wd0
                                    \raggedright
                                    \pretolerance10000 \hbadness10000
                                    \noindent \SSTitlefont
                                    \GetParenDim{\dimen0}{\dimen1}%
                                    \advance\dimen0 by\dimen1
                                    \baselineskip=\dimen0 #1}}
                    \hrule height0pt depth0pt
                    \dimen0=.8\baselineskip \advance\dimen0 by -2\parskip
                    \nobreak\vskip\dimen0 plus1pt minus1pt \noindent}%
\def\SubSectionvar #1\par{\Interskip{15pt}{3pt}{1pt}%
                    \leftline{\vbox{\noindent
                                    \SSTitlefont #1}}
                    \hrule height0pt depth0pt
                    \dimen0=.8\baselineskip \advance\dimen0 by -2\parskip
                    \nobreak\vskip\dimen0 plus 1pt minus1pt \noindent}%
\def\(#1){~\ifRomanNum {\Medrm\uppercase\expandafter{\romannumeral #1}}%
           \else {#1}%
           \fi}


\newcount\EqNo \EqNo=0
\def\NumbEq {\global\advance\EqNo by 1
             \eqno(\the\EqNo)}
\def\PrevEq {(\the\EqNo)}
\def\PrevEqs #1{{\count255=\EqNo \advance\count255 by-#1\relax
                 (\the\count255)}}
\def\NameEq #1{\xdef#1{(\the\EqNo)}}

\def\AppndEq #1{\EqNo=0
  \def\NumbEq {\global\advance\EqNo by 1
             \eqno(#1\the\EqNo)}
  \def\numbeq {\global\advance\EqNo by 1
             (#1\the\EqNo)}
  \def\PrevEq {(#1\the\EqNo)}
  \def\PrevEqs ##1{{\count255=\EqNo \advance\count255 by-##1\relax
                 (#1\the\count255)}}
  \def\NameEq ##1{\xdef##1{(#1\the\EqNo)}}}


\newcount\ThmNo \ThmNo=0
\newcount\LemNo \LemNo=0
\def\Theorem #1\par{\removelastskip\bigbreak
    \advance\ThmNo by 1
    \noindent{\bf Theorem \the\ThmNo:} {\sl #1\bigskip}}
\def\Lemma #1\par{\removelastskip\bigbreak
    \advance\LemNo by 1
    \noindent{\bf Lemma \the\LemNo:} {\sl #1\bigskip}}

\def\PrevThm {Theorem~\the\ThmNo}
\def\Thm {{\advance\ThmNo by 1 \PrevThm}}
\def\PrevLm {lemma~\the\LemNo}
\def\Declare #1#2\par{\removelastskip\bigbreak
    \noindent{\bf #1:} {\sl #2\bigskip}}
\def\Proof {\removelastskip\bigskip
    \noindent{\bf Proof:\ \thinspace}}
\def\EndProof{{~\nobreak\hfil \copy\Tombstone
               \parfillskip=0pt \bigskip}}


\newlinechar=`\!
\def\\{\ifhmode\hfil\break\fi}
\let\thsp=\,
\def\,{\ifmmode\thsp\else,\thinspace\fi}
\def\Hunskip {\ifhmode\unskip\fi}

\def\Date {\ifcase\month\or January\or February\or March\or April\or
 May\or June\or July\or August\or September\or October\or November\or
 December\fi\ \number\day, \number\year}

\newcount\mins  \newcount\hours
\def\Time{\hours=\time \mins=\time
     \divide\hours by60 \multiply\hours by60 \advance\mins by-\hours
     \divide\hours by60         
     \ifnum\hours=12 12:\ifnum\mins<10 0\fi\number\mins~P.M.\else
       \ifnum\hours>12 \advance\hours by-12 
         \number\hours:\ifnum\mins<10 0\fi\number\mins~P.M.\else
            \ifnum\hours=0 \hours=12 \fi
         \number\hours:\ifnum\mins<10 0\fi\number\mins~A.M.\fi
     \fi }

\def\HollowBox #1#2#3{{\dimen0=#1
       \advance\dimen0 by -#3 \dimen1=\dimen0 \advance\dimen1 by -#3
        \vrule height #1 depth #3 width #3
        \hskip -#3
        \vrule height 0pt depth #3 width #2
        \llap{\vrule height #1 depth -\dimen0 width #2}%
       \hskip -#3
       \vrule height #1 depth #3 width #3}}
\newbox\Tombstone
\setbox\Tombstone=\vbox{\hbox{\HollowBox{8pt}{5pt}{.8pt}}}

\def\GetParenDim #1#2{\setbox1=\hbox{(}%
                      #1=\ht1 \advance#1 by 1pt
                      #2=\dp1 \advance#2 by 1pt}

\def\Bull #1#2\par{\removelastskip\smallbreak\textindent{$\bullet$}
                    {\it #1}#2\smallskip}


\def\Title #1\par{\message{!!#1!}%
                  \nopagenumbers \pageno=-1
                  {\leftskip=0pt \rightskip=0pt
                   \Titlefont
                   \ifdim\baselineskip<2.5ex \baselineskip=2.5ex \fi
                   \indent #1\par}\bigskip\bigskip\bigskip}
\def\Author #1\par{{\count1=0   
                    \count2=0   
                    \dimen0=0pt
                    \Position{\ignorespaces\indent
                              #1\hskip-\dimen0 }\bigskip}}
\def\Address #1\par{{\leftskip=\parindent \rightskip=0pt
                     \parindent=0pt
                     \ifdim\baselineskip>3ex \baselineskip=2.7ex \fi
                     \sl #1\bigskip}}
\def\modfnote #1#2{{\parindent=1.1em \leftskip=0pt \rightskip=0pt
                    \GetParenDim{\dimen1}{\dimen2}%
                    \setbox0=\hbox{\vrule height\dimen1 depth\dimen2
                                          width 0pt}%
                    \advance\dimen1 by \dimen2 \baselineskip=\dimen1
                    \vfootnote{#1}{\hangindent=\parindent \hangafter=1
                                 \unhcopy0 #2\unhbox0
                                 \vskip-\gapskip }}}%
\def\PAddress #1{\ifcase\count1 \let\symbol=\dag
                 \or \let\symbol=\ddag
                 \or \let\symbol=\P
                 \or \let\symbol=\S
                 \else \advance\count1 by -3
                       \def\symbol{\dag_{\the\count1 }}%
                 \fi
                 \advance\count1 by 1
                 \setbox0=\hbox{$^{\symbol}$}\advance\dimen0 by .5\wd0
                 \Hunskip \box0
                 \modfnote{$\symbol$}{{\sl Permanent address\/}: #1}}%
\def\Email #1{\ifcase\count2 \let\symbol=\ast
                 \or \let\symbol=\star
                 \or \let\symbol=\bullet
                 \or \let\symbol=\diamond
                 \or \let\symbol=\circ
                 \else \advance\count2 by -4
                       \def\symbol{\ast_{\the\count2 }}%
                 \fi
                 \advance\count2 by 1
                 \setbox0=\hbox{$^{\textstyle\symbol}$}\advance\dimen0 by .5\wd0
                 \Hunskip \box0
                 \modfnote{$\symbol$}{{\sl Electronic mail\/}: #1}}%
\def\Abstract{\vfil \message{!Abstract:!}%
              \dimen0=\hsize \advance \dimen0 by-2\parindent
              \setbox0=\vbox\bgroup\hsize=\dimen0
              \Position{{\STitlefont Abstract}}
              \smallskip
              \ifdim\baselineskip>4.5ex \baselineskip=4.5ex \fi
              \noindent\ignorespaces}%
\def\EndAbstract{\par\egroup\noindent\hfil\shade{\copy0 }\par}
\def\shade #1{{\setbox1=\hbox{#1}\dimen0=\wd1
               \dimen1=\ht1 \advance\dimen1 by \dp1
               \advance\dimen0 by30pt
               \count255=\dimen0 \divide\count255 by 65536
               \advance\dimen1 by 10pt
               \count1=\dimen1 \divide\count1 by 65536
               \gray{\the\count255 }{\the\count1 }\hskip-\dimen0 \hskip14pt #1}}
\def\gray #1#2{\hbox to #1bp{
                                 \hss}}

\def\StartPaper{\pageno=1 \ifPubSub \footline{\tenrm\hss
                                              --\ \folio\ -- \hss}%
                          \else \footline{\hss\vtop to 0pt{\hsize=.15\hsize
                                     \vglue6pt \hrule \medskip
                                     \centerline{\tenrm\folio}\vss}\hss}%
                          \fi}

\newskip\gapskip \gapskip=\baselineskip


\RomanNumtrue
\let\Position=\leftline 
\def\TitleTypefaces #1#2#3{\font\Titlefont=#1
                         \font\STitlefont=#2
                         \font\SSTitlefont=#3\relax}
\def\Titlefont{\tenbf}   
\def\STitlefont{\tenbf}  
\def\SSTitlefont{\tensl} 
\def\Smallrm{\sevenrm}   
\def\Medrm{\tenrm}       
\newif\ifTR \TRfalse

\def\TimesRoman{\TRtrue
                \magnification=1050 \hoffset=0pt \voffset=0pt
                \input mtmacs2 
                \input 9pt-tms
                \pretolerance=800 \tolerance=1000
                \vbadness=1000 \hbadness=1000 \hfuzz=1 true pt
                \baselineskip=1.01\baselineskip
                \parskip=0pt plus 1pt
                \setbox\strutbox=\hbox{\vrule height .75\baselineskip
                              depth  .25\baselineskip
                              width 0pt}%
                \Page{4.75 true in}{6.875 true in}
                \def\Medrm{\Ninerm}
                \def\FootCapFace{\Ninepoint}
                \TitleTypefaces{pagd at 11pt}{pagd}{ptmro}
                \SmallFootnotes}


\def\Condition{limited influence condition}


\message{!<*><*><*><*><*><*><*><*><*><*><*><*><*><*><*><*><*><*><*><*><*><*><*>}
\message{!PROCESSING NOTE:!
         If you print the dvi file via `dvips' the figures will!
         automatically emerge. If you use anything else, you will!
         get the text of the paper, but no figures.!
         Send e-mail to borde@cosmos2.phy.tufts.edu if you have any!
         problems.!!}
\message{<*><*><*><*><*><*><*><*><*><*><*><*><*><*><*><*><*><*><*><*><*><*><*>!}

\ifTR
\else

\hoffset=0pt \voffset=0pt  
\pretolerance=800 \tolerance=1000
\vbadness=1000 \hbadness=1000 \hfuzz=1 true pt
\baselineskip=1.01\baselineskip
\parskip=0pt plus 1pt
\setbox\strutbox=\hbox{\vrule height .75\baselineskip
                              depth  .25\baselineskip
                              width 0pt}%

\Page{4.75 true in}{6.875 true in}


\TitleTypefaces{cmssdc10 at 12pt}{cmssdc10}{cmss10}


\def\LoadNine {\font\Ninerm=cmr9  \font\Ninesl=cmsl9
               \font\Nineit=cmti9 \font\Ninebf=cmbx9
               \font\Ninett=cmtt9 \font\Ninei=cmmi9
               \font\Ninesy=cmsy9
               \def\LoadNine{}}%

\def\Ninepoint {\def\rm{\fam0\Ninerm}%
  \textfont0=\Ninerm \scriptfont0=\sevenrm \scriptscriptfont0=\fiverm
  \textfont1=\Ninei  \scriptfont1=\seveni  \scriptscriptfont1=\fivei
  \textfont2=\Ninesy \scriptfont2=\sevensy \scriptscriptfont2=\fivesy
  \textfont3=\tenex   \scriptfont3=\tenex \scriptscriptfont3=\tenex
  \textfont\itfam=\Nineit  \def\it{\fam\itfam\Nineit}%
  \textfont\slfam=\Ninesl  \def\sl{\fam\slfam\Ninesl}%
  \textfont\ttfam=\Ninett  \def\tt{\fam\ttfam\Ninett}%
  \textfont\bffam=\Ninebf  \scriptfont\bffam=\sevenbf
     \scriptfont\bffam=\fivebf  \def\bf{\fam\bffam\Ninebf}%
  \normalbaselineskip=11pt
  \setbox\strutbox=\hbox{\vrule height8pt depth3pt width0pt}%
  \abovedisplayskip 9pt plus2pt  minus7pt
  \abovedisplayshortskip 0pt plus2pt
  \belowdisplayskip 9pt plus2pt  minus7pt
  \belowdisplayshortskip 6pt plus2pt  minus3pt
  \normalbaselines \rm }

\def\FootCapFace{\Ninepoint} 

\let\FFFFFF=\footstrut  
\def\SmallFootnotes {\LoadNine\def\footstrut{\Ninepoint\FFFFFF}}%

\SmallFootnotes

\fi 
\VersionInfo{To appear in the Proceedings of the Sixth Quantum Gravity
             Seminar, Moscow.}

\Title Singularities in Inflationary Cosmology: A Review

\Author Arvind Borde
\PAddress{Long Island University, Southampton, NY~11968.}
\Email{borde@cosmos2.phy.tufts.edu}
and Alexander Vilenkin
\Email{vilenkin@cosmos2.phy.tufts.edu}

\Address Institute of Cosmology\\
         Department of Physics and Astronomy\\
         Tufts University\\
         Medford, MA 02155, USA.

\Sectionvar \indent

{\leftskip=\parindent \rightskip=\parindent \Ninepoint
{\bf Abstract:}
We review here some recent results that show that inflationary
cosmological models must contain initial singularities.
We also present a new singularity theorem.
The question of the initial singularity re-emerges in
inflationary cosmology
because inflation is known to be generically future-eternal.
It is natural to ask, therefore, if inflationary models can be
continued into the infinite past in a non-singular way.
The results that we discuss show that the answer to the
question is ``no.'' This means that we cannot use
inflation as a way of avoiding the question of the birth
of the Universe. We also argue that our new theorem
suggests~-- in a sense that we explain in the
paper~-- that the Universe cannot be infinitely old.
\par}

\StartPaper

\Section Introduction

Inflationary cosmological models appear, at first glance, to admit the
possibility that the Universe might be described by a version
of the steady-state picture. The possibility seems to arise because
inflation is generically future-eternal: in a large class of
inflationary cosmological models
the Universe consists of a number of isolated thermalized regions
embedded in an always-inflating background
\Ref{The inflationary expansion is driven by the potential energy of a
     scalar field $\varphi$, while the field slowly ``rolls down'' its
     potential $V(\varphi)$. When $\varphi$ reaches the minimum of the
     potential this vacuum energy thermalizes, and inflation is followed by
     the usual radiation-dominated expansion. The evolution of the field
     $\varphi$ is influenced by quantum fluctuations, and as a result
     thermalization does not occur simultaneously in different parts of the
     Universe.}.
The boundaries of the thermalized regions expand into this background,
but the inflating domains that
separate them expand even faster, and the thermalized regions do not,
in general, merge.
As previously created regions expand, new ones come into existence,
but the Universe does not fill up entirely with thermalized regions
\Refl{A. Vilenkin, \Jou{Phys. Rev.~D}, \Vol{27}, 2848 (1983);\\
      A.D. Linde, \Jou{Phys. Lett. B\/} \Vol{175}, 395 (1986).}
\StoreRef{\AryaVi}
\Refn{M. Aryal and A. Vilenkin, \Jou{Phys. Lett. B\/} \Vol{199}, 351 (1987);\\
      A.S. Goncharov, A.D. Linde and V.F. Mukhanov,
        \Jou{Int. J. Mod. Phys. A\/} \Vol{2}, 561 (1987);\\
      K. Nakao, Y. Nambu and M. Sasaki, \Jou{Prog. Theor. Phys.\/}
        \Vol{80}, 1041 (1988).}
\StoreRef{\Linde}
\Refe{A. Linde, D. Linde and A. Mezhlumian, \Jou{Phys. Rev.~D}, \Vol{49},
        1783 (1994).}.
A cosmological model in which the inflationary phase has no global
end and continually produces new ``islands of thermalization''
naturally leads to this question:
can the model be extended in a non-singular way into the infinite past,
avoiding in this way the problem of the initial singularity?
The Universe would then be
in a steady state of eternal inflation without a beginning.

Assuming that some rather
general conditions are met, we have recently shown
\StoreRef{\BVOne}
\Refl{A. Borde and A. Vilenkin, \Jou{Phys. Rev. Lett.}, \Vol{72}, 3305
      (1994).}
\StoreRef{\BOne}
\Refn{A. Borde, \Jou{Phys. Rev. D.}, \Vol{50}, 3392 (1994).}
\StoreRef{\BVTwo}
\Refn{A. Borde and A. Vilenkin, in \Jou{Relativistic Astrophysics:
      The Proceedings
      of the Eighth Yukawa Symposium}, edited by M.\ Sasaki,
      Universal Academy Press, Japan (1995).}
\StoreRef{\BTwo}
\Refe{A. Borde, Tufts Institute of Cosmology preprint (1995).}
that the answer to this question is ``no'':
generic inflationary models necessarily contain initial singularities.
This is significant, for it forces us
in inflationary cosmologies (as in the standard big-bang ones)
to face the question of what, if anything, came before.

This paper reviews what is known about the existence of
singularities in inflationary cosmology.
A partial answer to the singularity question was previously given
by Vilenkin
\StoreRef{\Vilenkin}
\Ref{A. Vilenkin, \Jou{Phys. Rev.~D}, \Vol{46}, 2355 (1992).}
who
showed the necessity of a beginning in a two-dimensional spacetime
and gave a plausibility argument for four dimensions. The
broad question was also previously addressed by Borde
\StoreRef{\Borde}
\Ref{A. Borde, \Jou{Cl. and Quant. Gravity\/} \Vol{4}, 343 (1987).}
who sketched
a general proof using the Penrose-Hawking-Geroch global techniques.
We will not discuss this earlier work here, concentrating instead on
more recent results.

The paper is organized as follows:
Section~II outlines some mathematical background
(see Hawking and Ellis
\StoreRef{\HE}
\Ref{S.W. Hawking and G.F.R. Ellis, \Jou{The large scale structure of
      spacetime}, Cambridge University Press, Cambridge, England (1973).}
for details).
Section~III describes our first theorem, applicable to open
Universes with a simple causal structure.
Section~IV sketches how the theorem may be extended to
closed Universes.
Section~V presents a new theorem:
Here, we drop the assumption that the causal structure of the
Universe is simple.
Instead, we introduce a new condition, which we call the
{\it \Condition}. We argue that this condition is
likely to hold in many
inflationary models. Our new theorem
makes no assumptions about whether the Universe is open or closed,
thus providing a unified treatment of the two cases.
Section~VI offers some concluding comments.

\Section Mathematical Preliminaries

Spacetime is represented by a manifold~$\cal M$ with a
{\it time-oriented\/}
\Ref{This means that the notions of ``past'' and ``future''
     are globally well-defined.}
Lorentz metric $g_{ab}$ of signature $(-, +, +, +)$. We do not
assume any specific field equation for~$g_{ab}$. Instead,
we impose an inequality on the Ricci curvature
(called a {\it convergence condition\/}),
and our conclusions are valid in any theory
of gravity (such as Einstein's, with a physically reasonable source)
in which such a condition is satisfied.

A curve is called {\it causal\/} if it is everywhere either timelike
or null.
The {\it causal\/} and {\it chronological pasts\/} of a
point~$p$, denoted
respectively by $J^-(p)$ and $I^-(p)$, are defined as follows:
\medskip

$J^-(p) = \{q:$ $\exists$ a future-directed causal curve
           from $q$ to $p\}$,

\smallskip
\noindent and
\smallskip

$I^-(p) = \{q:$ $\exists$ a future-directed timelike curve
           from $q$ to $p\}$.

\medskip\noindent
The futures $J^+(p)$ and $I^+(p)$ are defined similarly.
The sets $I^{\pm}(p)$ are open: i.e., if $x\in I^{\pm}(p)$, then
all points in some neighborhood of~$x$ also lie in~$I^{\pm}(p)$.
The {\it past light cone\/} of~$p$ is defined~[\BOne]
as~$E^-(p) = J^-(p) - I^-(p)$. It follows that $E^-(p)$
is {\it achronal\/} (i.e., no two points on it can be connected by
a timelike curve) and that
$E^-(p) \subset \dot I^-(p)$ (where $\dot I^-(p)$ is the boundary of
$I^-(p)$).
In general, however, $E^-(p) \ne \dot I^-(p)$ (see~\NFig).
These definitions of futures, pasts, and light cones can be
extended from single points~$p$ to arbitrary spacetime sets
in a straightforward manner.

Spacetimes in which
$E^-(p) = \dot I^-(p)$, for all points~$p$, are called {\it past causally
simple}. We tighten this definition by further requiring that
$E^-(p) \ne \emptyset$ (this rules out certain causality violations).
\StartFigure{190}{115}
            {newpath 0 0 moveto
             90 90 lineto
             140 40 lineto
             170 70 lineto
             stroke
             newpath 0 0 moveto  
             90 90 lineto
             140 40 lineto
             170 70 lineto
             190 70 lineto
             210 50 170 20 190 0 curveto
             130 -20 60 20 0 0 curveto
             gsave .9 setgray fill grestore
             2 setlinewidth
             .3 setgray
             newpath 80 70 moveto
             100 70 lineto
             stroke
             newpath 170 70 moveto
             190 70 lineto
             stroke
             .6 setlinewidth
             0 setgray
             gsave
             [3 3] 0 setdash
             newpath 160 45 moveto
             175 68 lineto
             stroke
             newpath 85 72 moveto
             90 89 lineto
             stroke
             /solidcirc { newpath
             1 0 360 arc gsave 0 setgray fill grestore} def
             90 90 solidcirc
             160 45 solidcirc
             grestore
             .3 setlinewidth
             /arrow { newpath 0 0 moveto 0 -5 lineto 2 -2 lineto
                      0 -5 moveto -2 -2 lineto stroke newpath 0 0 moveto} def
             gsave 50 58 translate arrow grestore
             gsave 115 73 translate arrow grestore
             gsave 155 63 translate arrow grestore
             newpath 10 107 moveto
             115 107 115 73 15 arcto
             115 73 lineto stroke
             newpath 10 105 moveto
             50 105 50 58 20 arcto
             50 58 lineto stroke
             newpath 167 95 moveto
             155 95 155 63 10 arcto
             155 63 lineto stroke
          }
\label(93,94){$p$}
\label(164,45){$q$}
\label(-20,104.5){E$^-(p)$}
\label(171,92){$\dot {\rm I}^-(p)-{\rm E}^-(p) \ne \emptyset$}
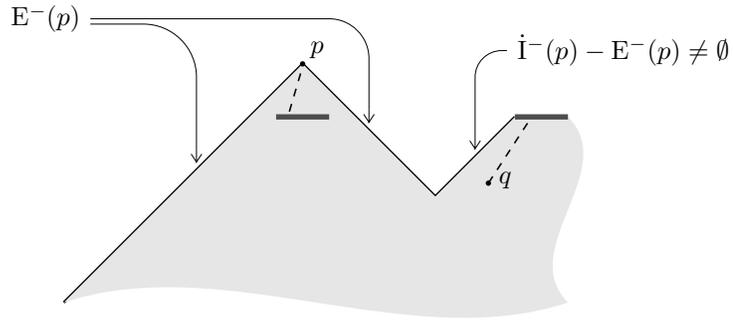
\caption{An example of the causal complications that can arise in an
unrestricted spacetime. Light rays travel along 45$^\circ$ lines in this
diagram, and the two thick horizontal lines are identified. This allows
the point~$q$ to send a signal to the point~$p$ along the dashed line,
as shown, even though~$q$ lies outside what is usually considered
the past light cone of~$p$. The boundary of the past of~$p$,
$\dot {\rm I}^-(p)$, then consists of the past light cone of~$p$,
$E^-(p)$, plus a further piece.
Such a spacetime is not ``causally simple.''}
\EndFigure

A timelike curve is {\it maximally extended\/} in the past direction
if it has no past endpoint. (Such a curve is often called
past-inextendible.) The idea behind this is that such a curve
is fully extended in the past direction, and is not merely a segment
of some other curve.

We define a {\it closed Universe\/} as one that
contains a compact, edgeless, achronal hypersurface,
and an {\it open Universe\/} as one that contains no such surface.

The {\it strong causality condition\/} holds on~$\cal M$ if
there are no closed or ``almost-closed'' timelike or
null curves through any point of~$\cal M$. If $\mu$ is any timelike
curve in a spacetime that obeys the strong causality condition
and~$x$ is any point not on~$\mu$, then there must be some neighborhood
$\cal N$ of~$x$ that does not intersect~$\mu$. (Otherwise, $\mu$
would accumulate at~$x$, and thereby give an almost-closed
timelike curve.)

Finally, consider a congruence
\Ref{A congruence is a set of curves in an open region of spacetime,
one through each point of the region.}
of null geodesics with affine parameter~$v$
and tangent~$V^a$. The expansion of the geodesics may be defined
as $\theta\equiv D_a V^a$, where $D_a$ is the covariant derivative.
The propagation
equation for $\theta$ leads to this inequality:
$$
{d\theta\over dv} \leq -{1\over 2}\theta^2 - R_{ab}V^aV^b\; .\NumbEq
$$
Suppose that (i)~$R_{ab}V^aV^b\geq0$ for all null vectors~$V^a$
(this is called the {\it null convergence condition\/}),
(ii)~the expansion,~$\theta$, is negative at some
point~$v=v_0$ on a geodesic~$\gamma$,
and (iii)~$\gamma$ is complete in the direction of
increasing~$v$ (i.e., $\gamma$ is defined for all~$v \geq v_0$).
Then $\theta \to -\infty$ along~$\gamma$ a finite affine parameter
distance from~$v_0$~[\HE\,
\Refe{A weakening of the conditions under which $\theta$ diverges,
     was discussed by\\
     F.J. Tipler, \Jou{J. Diff. Eq.}, {\bf 30}, 165 (1978);
      \Jou{Phys. Rev.~D}, {\bf 17}, 2521 (1978);\\
     these results were extended in~[\Borde].}.

\Section Open Universes

Our first result~[\BVOne\,\BVTwo] applies to open, causally simple
spacetimes:

\Theorem
A spacetime~$\cal M$ cannot be null-geodesically complete to the past
if it satisfies the following conditions:
\item{A.} It is past causally simple.
\item{B.} It is open.
\item{C.} It obeys the null convergence condition.
\item{D.} It has at least one point~$p$ such that for
every point~$q$ to the past of~$p$ the
volume of the difference of the pasts of~$p$ and~$q$ is finite.

Assumptions~A--C are conventional as far as work on singularity theorems
goes. But assumption~D is new and is inflation-specific.
A slightly different version
has been discussed in detail elsewhere~[\Vilenkin\,\BVTwo],
but here is a rough, short explanation:
It may be shown that if a point~$r$
lies in a thermalized region, then all points in~$I^+(r)$ also lie in
that thermalized region~[\BVOne]. Therefore,
given a point~$p$ in the inflating region, all points in its
past must lie in the inflating region.
Further, it seems plausible that
there is a zero probability for no thermalized regions to form in an
infinite spacetime volume. Then assumption~D follows.

\Proof
The full proof of this result is available elsewhere~[\BVOne\,\BVTwo],
but here is a sketch:
Suppose that~$\cal M$ is null-complete to the past.
We show that a contradiction follows.

Let $q$ be a point to the past of the point~$p$ of assumption~D.
Then every past-directed null geodesic from~$q$ must leave
$E^-(q)$ at some point and enter $I^-(q)$ (i.e., it must leave the
past null cone of~$q$ and enter the interior of the past of~$q$).
For, let $\gamma$ be a  past-directed null geodesic from~$q$, and
suppose that $\gamma$ lies in $E^-(q)$ throughout.
Choose a small ``triangle'' of null geodesics
neighboring~$\gamma$ in $E^-(q)$ and construct a volume ``wedge''
by moving the triangle so that its vertex moves from~$q$ to
a point $q'$ (still in $I^-(p)$), an infinitesimal distance to
the future of~$q$.
The volume of this region may be expressed~[\BVOne\,\BVTwo] as
$$
\Delta\int_0^\infty\!\! {\cal A}(v) \,dv,
$$
where $\Delta$ is a constant, $\cal A$~is the cross-sectional area
of~$E^-(q)$ in the wedge, and $v$~is an affine parameter along
the geodesic (chosen to increase in the past direction).
From assumption~D, this volume (being a part
of the volume of $I^-(p)-I^-(q)$) must be finite.
This can happen only if~$\cal A$ decreases somewhere. But
$$
{d{\cal A}\over dv} = \theta {\cal A},
$$
where~$\theta$ is the divergence of the congruence of
null geodesics that make up our volume wedge.
Therefore, $\theta$ must become negative somewhere.
We have seen that assumption~C then implies that $\theta\to -\infty$
within a finite affine parameter distance.
It follows from this (by a standard argument~[\HE])
that~$\gamma$ must leave $E^-(q)$ and enter $I^-(q)$,
a finite affine parameter distance from~$q$.

Now, this holds for any geodesic~$\gamma$ that lies in
$E^-(q)$ sufficiently far to the past.
Thus~$E^-(q)$ is compact. But assumption~A implies
that~$E^-(q)$ has no edge. These two statements taken together
contradict assumption~B.
\EndProof

\Section Closed Universes

Closed Universes are potentially awkward for our theorem because
it is possible for light cones in some closed Universes
to ``wrap around'' the whole Universe, and thus be compact,
without causing any problems. This is illustrated in \NFig
\Ref{Similar behavior occurs, for instance, in the Einstein Universe,
     but it does not in the de~Sitter Universe, nor in some closed
     Robertson-Walker Universes~[\HE\,\BOne].}.
Such behavior is unreasonable, however, in the context of most
inflationary cosmological models,
which are ``spatially large'' in the sense that they
contain many different regions that are not in causal communication.
\StartFigure{92}{210}
   {
    gsave
    1.5 setlinewidth
    1 .25 scale
    newpath 46 92 46 0 360 arc stroke
    newpath 46 748 46 0 180 arc stroke
    newpath 46 748 46 180 0 arc gsave [2 2] 0 setdash stroke grestore
    grestore
    newpath 0 23 moveto 0 187 lineto stroke
    newpath 92 23 moveto 92 187 lineto stroke
    newpath 20 150 moveto 15 148 5 138 0 130 curveto stroke
    newpath 20 150 moveto 40 140 72 108 92 78 curveto stroke
    gsave [2 2] 0 setdash
    newpath 0 120 moveto 20 90 52 63 72 58 curveto stroke
    newpath 92 73 moveto 87 65 77 60 72 58 curveto stroke
    grestore
    newpath 20 150 1.5 0 360 arc fill
    newpath 72 58 1.5 0 360 arc fill
   }
\label(23,153){$q$}
\label(75,54){$r$}
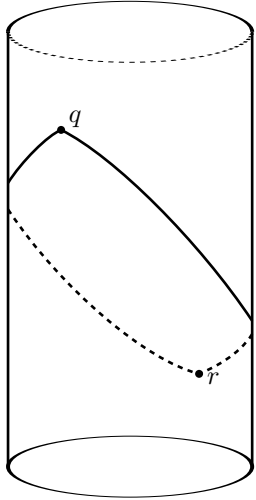
\caption{A closed Universe in which the past light cone
of any point~$q$ is compact (and the volume of the difference of the
pasts of any two points is finite). The past-directed null geodesics
from~$q$ start off initially in~$E^-(q)$; but, once they recross
at~$r$ (``at the back'') they enter~$I^-(q)$ (because
there are timelike curves between~$q$ and points on these null geodesics
past~$r$), and they thus leave~$E^-(q)$.}
\EndFigure

We define a {\it localized\/} light cone as one that does not wrap
around the Universe. More precisely,
we say that a past light cone
is localized if from every spacetime
point~$p$ not on the cone there is at least
one timelike curve, maximally extended in the past direction,
that does not intersect the cone
\Ref{We actually need to impose a further causality requirement,
     called the {\it stable causality\/} condition, in order
     for this definition be meaningful; see ref.~[\BOne]
     for the details.}.
It turns out that the conclusion of our theorem still
holds if we replace assumption~B by
the assumption that past light cones are
localized~[\BOne].

\Section Causally Complicated Universes

The assumption of causal simplicity~-- made in our first
result in order to simplify the proof~-- can be dropped,
as long as we are willing to make a replacement assumption
about the causal structure of inflating spacetimes. The
new theorem embraces topologically and causally complicated
spacetimes, and it allows us to give a unified treatment
of open and closed Universes.

\Theorem
A spacetime~$\cal M$ cannot be null-geodesically complete to the past
if it satisfies the following conditions:
\item{A.} It obeys the null convergence condition.
\item{B.} It obeys the strong causality condition.
\item{C.} It has at least one point~$p$ such that
\itemitem{i.} for every point~$q$ to the past of~$p$ the
              volume of the difference of the pasts of~$p$
              and~$q$ is finite (i.e., $\Omega(I^-(p)-I^-(q))<\infty$),
              and
\itemitem{ii.} there is a timelike curve~$\mu$,
               maximally extended to the past of~$p$, such that
               the boundary of the future of~$\mu$ has a non-empty
               intersection with the past of~$p$ (i.e.,
               $\dot I^+(\mu)\cap I^-(p) \ne \emptyset$).

Part~(ii) of assumption~C is new.
It is related to
certain other causal and topological properties of
spacetimes~[\BTwo],
and there are also physical reasons for believing that
the assumption is reasonable. Consider, for instance,
a point~$r$ in the inflating region. Suppose that its past,
$I^-(r)$, has the property that it ``swallows the Universe,''
in the sense that every timelike curve that is maximally
extended in the past direction eventually enters~$I^-(r)$.
(This is related to the issue of localization of light cones
discussed above.) Assuming that there are thermalization
events arbitrarily far in the past, it seems likely, then, that
there is a thermalization event somewhere in~$I^-(r)$. This
contradicts the fact that $r$~lies in the inflating
region~[\BVOne].
It is plausible, therefore, that inflating spacetimes will,
in general, have the property that
there exist maximally extended (in the past direction) timelike
curves whose futures do not encompass the whole inflating region.
(If no timelike curve has a future that encompasses the entire
inflating region, it will guarantee that the Universe never
completely thermalizes~-- so one may view a condition of this
sort as a sufficient condition for inflation to be future-eternal.)

Another piece of evidence for the reasonableness of
part~(ii) of assumption~C is that
the spacetime in the past light cone of any point in the inflating
region is locally approximately de Sitter. It is similar to the
spacetime in the future light cone of a point in
an inflating universe where
there is no thermalization. Thus ``past infinity'' in inflating
regions might be expected to be similar to that of de~Sitter space,
where the sort of behavior we are talking about does occur~[\HE].

We are arguing, in other words, that a typical maximally extended
past-directed curve ought not to influence the entire
inflating region~--
there must be portions of the region that do not
lie to the future of such a curve. This is illustrated in~\NFig.
Let~$\cal V$ be a spacetime region.
We call a timelike curve,~$\mu$,
a curve of limited influence in~$\cal V$ if its future does not
engulf all of~$\cal V$.
If~$\cal V$ is the inflating region of a
spacetime~$\cal M$, and if all timelike curves
in~$\cal M$ are of limited influence in~$\cal V$, we say that
the spacetime obeys the {\it \Condition}.

\StartFigure{290}{100}
    {
     gsave
     newpath 50 0 moveto 0 50 lineto 50 100 lineto 100 50 lineto
     closepath clip
     .5 setlinewidth
     newpath 50 0 moveto 0 50 lineto 50 100 lineto
     100 50 lineto closepath
     gsave .9 setgray fill grestore
     1 setlinewidth
     newpath 50 69 moveto
     [1 1] 0 setdash 45 50 50 20 50 0 curveto stroke
     grestore
     newpath 50 0 moveto 0 50 lineto 50 100 lineto 100 50 lineto
     closepath stroke 
     gsave
     .5 setlinewidth
     newpath 145 80 moveto 205 20 lineto 285 80 lineto
     gsave .9 setgray fill grestore stroke
     1 setlinewidth
     newpath 210 69 moveto
     [1 1] 0 setdash 205 55 205 35 205 20 curveto stroke
     grestore
     newpath 140 20 moveto 290 20 lineto 290 80 lineto
             140 80 lineto closepath stroke 
    }
\label(38,48){$\mu$}
\label(20,0){(a)}
\label(196,48){$\mu$}
\label(200,0){(b)}
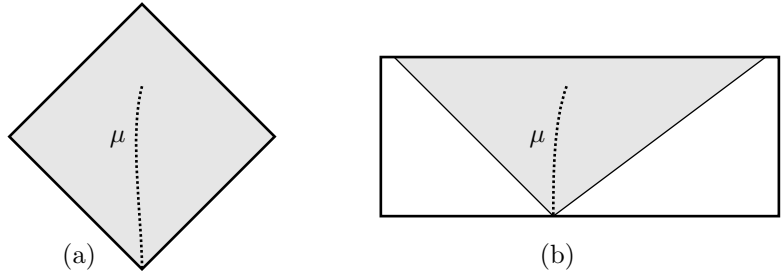
\caption{These figures each represent the inflating region of
some spacetime. The
shaded region in each case represents the future of the
curve~$\mu$. In~(a) $\mu$~can influence the entire inflating
region, whereas
in~(b) it cannot.}
\EndFigure

\Proof
Suppose that~$\cal M$ is null-complete to the past. We show that
this leads to a contradiction.

Let $q$ be a point to the past of the point~$p$ of assumption~C.
We have seen in Theorem~1 that
every past-directed null geodesic from~$q$ must leave
$E^-(q)$ at some point and enter $I^-(q)$ (i.e., it must leave the
past null cone of~$q$ and enter the interior of the past of~$q$).

\StartFigure{190}{115}
            {newpath 0 0 moveto
             90 90 lineto
             180 0 lineto
             stroke
             newpath 0 0 moveto  
             90 90 lineto
             180 0 lineto
             0 0 lineto
             gsave .9 setgray fill grestore
             /solidcirc { newpath
             1.5 0 360 arc gsave 0 setgray fill grestore} def
             /hollowcirc { newpath
             2 0 360 arc gsave 1 setgray fill grestore
             gsave 0 setgray .5 setlinewidth stroke grestore} def
             90 90 solidcirc
             100 0 hollowcirc
             120 20 solidcirc
             gsave  
             [1 1] 0 setdash
             .6 setlinewidth
             newpath 90 70 moveto
             80 50 110 30 100 2 curveto stroke
             grestore
             gsave  
             [2 1] 2 setdash
             .8 setlinewidth
             newpath 101 1 moveto
             145 45 lineto stroke
             newpath 99 1 moveto
             45 55 lineto stroke
             grestore
             .3 setlinewidth
             /arrow { newpath 0 0 moveto 0 -5 lineto 2 -2 lineto
                      0 -5 moveto -2 -2 lineto stroke newpath 0 0 moveto} def
             gsave 46 62 translate arrow grestore
             gsave 144 52 translate arrow grestore
             newpath 20 107 moveto
             144 107 144 52 15 arcto
             144 52 lineto stroke
             newpath 20 105 moveto
             46 105 46 62 15 arcto
             46 62 lineto stroke
          }
\label(93,94){$p$}
\label(80,60){$\mu$}
\label(123,17){$q$}
\label(113,5){$\gamma$}
\label(30,10){$I^-(p)$}
\label(-8,103){$\dot I^+(\mu)$}
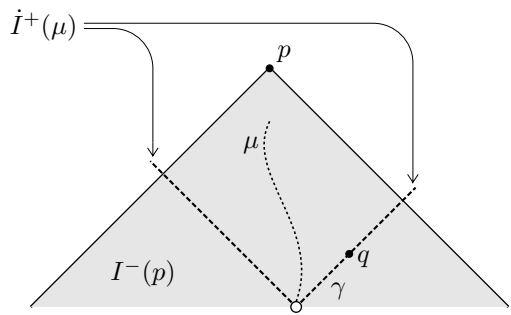
\caption{The null geodesic $\gamma$, past-directed from~$q$,
lies both on $\dot I^+(\mu)$ and on $E^-(q)$. It must lie on
$\dot I^+(\mu)$
throughout when followed into the past
(the hollow circle at the ``past end''
of~$\mu$ is not part of the spacetime). But~$q\in I^-(p)$,
so $\gamma$ must enter $I^-(q)$. This contradicts the
fact that it lies on $\dot I^+(\mu)$ throughout.}
\EndFigure
Let the point~$q$ belong to $\dot I^+(\mu)\cap I^-(p)$ (see~\Fig).
Let~$\gamma$ be
a null geodesic through~$q$ that lies on~$\dot I^+(\mu)$.
From assumption~B it follows that this
geodesic cannot leave $\dot I^+(\mu)$ when
followed in the past direction. For, suppose it does at some point~$x$.
This point cannot lie on~$\mu$ itself (because then it, and all points
to its causal future, including~$q$, will lie to the chronological
future of some point on~$\mu$, i.e., in $I^+(\mu)$ and not on its
boundary). Pick a neighborhood~$\cal N$ of~$x$ that does not intersect
$\mu$ anywhere (see the discussion of strong causality in Section~II).
There will be
some null geodesic in~$\cal N$, past-directed from~$x$,
that lies on
the boundary~$\dot I^+(\mu)$. If this geodesic, $\lambda$, is
other than the continuation of~$\gamma$, there will be a timelike
curve from it to a point on~$\gamma$ (see~\NFig), violating the
achronal nature of the boundary~$\dot I^+(\mu)$.

\StartFigure{100}{110}
            {/solidcirc { newpath
             1.5 0 360 arc gsave 0 setgray fill grestore} def
             newpath
             gsave
             2 setlinewidth 50 50 50 0 360 arc
             gsave .95 setgray fill grestore stroke grestore
             newpath 100 100 moveto 50 50 lineto stroke
             newpath 100 0 moveto 50 50 lineto stroke
             newpath 50 50 solidcirc
             gsave newpath
             [2 1] 2 setdash
             .8 setlinewidth
             70 70 moveto 65 56 75 44 70 30 curveto stroke
             grestore
             }
\label(20,65){$\cal N$}
\label(42,50){$x$}
\label(90,100){$\gamma$}
\label(87,0){$\lambda$}
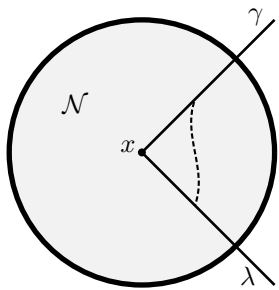
\caption{If the null geodesic $\gamma$ encounters at~$x$ some
other geodesic~$\lambda$, then there will be a timelike curve~--
shown by the dashed line~-- between the two.}
\EndFigure
Now, we have seen
that $\gamma$ must leave $E^-(q)$ and enter $I^-(q)$; i.e.,
there must be a point~$r$ to the past of~$q$ on~$\gamma$ such that
$r\in I^-(q)$. This means that every point in some neighborhood
of~$r$ must also lie in $I^-(q)$. Some of these points must belong
to~$I^+(\mu)$. (The point~$r$ lies on~$\gamma$, and so belongs
to~$\dot I^+(\mu)$, the boundary of the future of~$\mu$.
Therefore, there must
be points close to~$r$ that lie in~$I^+(\mu)$.) This means that there
is a timelike curve that starts in the past at some point on~$\mu$,
passes through a point close to~$r$, and then continues on to~$q$.
This contradicts the fact that $q\in \dot I^+(\mu)$.
\EndProof

\Section Discussion

The theorems in this paper show that inflation does not
seem to remove the
problem of the initial singularity (although it does move the
singularity back into an indefinite past).
In fact, our analysis of the assumptions of the theorems suggests
that almost all points in the inflating region have a singularity
somewhere in their pasts. In this sense, our results
are stronger than most of the usual singularity theorems,
which~-- in general~-- predict the existence of just one incomplete
geodesic
\Ref{Our results are also stronger than many standard singularity
     theorems~-- such as the Hawking-Penrose theorem~[\HE]~--
     because we do not assume the {\it strong energy condition}.
     This is crucially important when discussing the structure
     of inflationary spacetimes, because the condition is explicitly
     violated there~[\BVTwo].}.

Indeed, \PrevThm\ is even stronger than that, since it appears to
suggest that the Universe cannot be infinitely old, in the sense that
the inflating region of spacetime can contain no timelike curve 
infinitely long (in proper time) in the past direction
\StoreRef{\LLM}
\Ref{The existence, or not, of such a curve is related to issues raised
      in\\
      A.D. Linde, D. Linde and A. Mezhlumian, \Jou{Phys. Rev.~D},
     \Vol{49}, 1783 (1994).}.
For, suppose such a curve, $\mu$, does exist. 
It seems reasonable to suppose that
the null geodesics that lie on the boundary of the future of~$\mu$
are also complete in the past direction
\Ref{It is possible to contrive examples in which this is not true~--
     where, for instance, a timelike curve avoids singularities in the 
     past, but no null ones in the boundary of its future do. 
     In a physically reasonable spacetime, however, one
     would expect singularities to be visible to timelike and null
     curves alike.}. 
If this is the case, and if~$\mu$ is of limited influence in
the inflating region, we arrive at the same contradiction as the
one in our theorem
\Ref{The question of whether or not the Universe is infinitely old 
     is sometimes posed as the question of whether or not there exists 
     an upper bound to the length of timelike curves when followed into
     the past. This formulation does not, however, get to the 
     essence of the question. Consider, for example, two-dimensional 
     Minkowski space with the region $t\leq 0$ removed. This truncated 
     spacetime has a ``global beginning'' at $t=0$, and is thus not 
     infinitely old at any (finite) positive time~$t$.
     When viewed from the spacelike hypersurface, $\cal S$, 
     given by $t=\sqrt{1+x^2}$, however, the $t=0$ surface appears 
     further and further in the past the further one moves away
     from~$x=0$. There is no 
     upper bound here on the lengths of past-directed timelike curves 
     from~$\cal S$.}.

The existence of initial singularities in inflationary models means
that we cannot use inflation as a way of avoiding the question of
the birth of the Universe. The question will probably have to be answered
quantum mechanically, i.e., by describing the Universe by a wave function,
and not by a classical spacetime.

\Sectionvar Acknowledgements

One of the authors (A.V.) acknowledges partial support from the National
Science Foundation. The other author (A.B.) thanks the Institute of
Cosmology at Tufts University and Dean Al Siegel and Provost Tim Bishop
of Southampton College of Long Island University for their continued 
support.

\EndPaper